# Realizing Wardrop Equilbria with Real-Time Traffic Information

L. C. Davis, 10244 Normandy Dr., Plymouth, Michigan 48170

**Abstract**

A Wardrop equilibrium for multiple routes from the same origin to the same destination requires equal travel time on each path used. With the advent of real-time traffic data regarding travel times on alternative routes, it becomes important to analyze how best to use the information provided to drivers. In particular, can a Wardrop equilibrium, which is a desired state, be realized? Simulations using a realistic traffic model (the three-phase model) on a two-route example are presented to answer this question. One route (the main line) is a two-lane highway with a stalled vehicle in the right lane and the other route is a low-speed bypass. For a critical incoming flow of vehicles, a phase transition between free flow and congested flow near the stalled vehicle is observed, making this a challenging example. In the first scenario, drivers choose routes selfishly on the basis of current travel times. The result is strong oscillations in travel time because of the inherent delay in the information provided. The second scenario involves a hypothetical control system that limits the number of vehicles on the main line to prevent the free-flow to congested-flow phase transition by diverting sufficient flow to the bypass. The resulting steady state is neither a Wardrop equilibrium nor a system optimum, but an intermediate state in which the main-line travel time is less than on the bypass but the average for all vehicles is close to a minimum. In a third scenario, anticipation is used as a driver-advice system to provide a fair indicator of which route to take. Prediction is based on real-time data comparing the number of vehicles on the main line at the time a vehicle leaves the origin to the actual travel time when it reaches the destination. Steady states that approximate Wardrop equilbria, or at least as close to them as can be expected, are obtained. This approach is also applied to an example with a low-speed boundary condition imposed at the destination in place of a stalled vehicle. The steady state flow approaches a Wardrop equilibrium because there is no abrupt change in travel time due to a phase transition.

# I. Introduction

One of the important concepts in transportation theory is the Wardrop equilibrium. [1] For alternative routes connecting the same origin and destination, Wardrop stated in 1952 that

> "The journey times on all routes actually used are equal, and less than those which would be experienced by a single vehicle on any unused route."

When this condition is realized, the flow of traffic is in equilibrium.

Wardrop further postulated that the average travel time for all vehicles is a minimum at equilibrium, although this is now known to be only approximately correct. Recent research activity has focused on just how good (or bad) such equilibria are [2-4]. In other terms, selfish routing (drivers choosing the route with the lowest expected travel time) is an example of a non-cooperative game and the Wardrop equilibrium is a Nash equilibrium, which does not have to be a system optimum. [5]

Only a few studies have been reported in which realistic traffic-model simulations show the approach to equilibrium with selfish routing. Lo and Szeto [6] used the Cell Transmission Model [7,8] to demonstrate an ideal dynamic user optimum. Wahle *et al.* [9,10] used the Nagel-Schreckenberg cellular automaton model [11] on a two-route scenario to calculate the response to real-time traffic information. An equilibrium (the authors did not use the term "Wardrop equilibrium") did not develop; rather an oscillatory pattern in the travel time for each route was found due to the delayed nature of the information (travel times are only as recent as the last car to complete the route). Because the two routes in this example are equivalent, a Wardrop equilibrium would require equal, constant travel times. Instead, out-of-phase variations of travel times were obtained.

With the continuing improvement in real-time information for drivers regarding alternative routes [10, 12-15], it is pertinent to find appropriate algorithms to make maximum use of the information. Many studies of the Wardrop equilibrium are based on historical data and previous experiences of drivers to find the optimum solution iteratively. (See Refs. [16], [17], [18], and [19] and references therein.) In contrast, the focus of the present paper is more closely related to previous research when advances in communication were predicted and their potential consequences were considered. [20-23]

In this paper, an inequivalent two-route model is analyzed using the three-phase traffic model of Kerner and Klenov [24-28]. The main route is a two-lane freeway with a stalled vehicle (or any local lane blockage) in the right lane. The other route is a bypass that allows vehicles from the right lane to exit prior to encountering a stalled vehicle. Three scenarios are considered: (1) selfish routing where drivers have complete travel-time information available and can decide to take the bypass if it has a shorter travel time; (2) a hypothetical control system in which drivers are directed to exit onto the bypass based on the number of vehicles on one or both paths; and (3) an advisory system where the anticipated better route (with the shorter travel time) is broadcast to drivers. The purpose of the analysis is to determine if a Wardrop equilibrium can be attained in a realistic traffic simulation with complete and up-to-date information and to evaluate alternative routing algorithms to improve travel times. Although the three-phase model has been criticized [29], it is sufficiently accurate for the present analysis.

Congestion induced by a blockage on one lane of a two-lane highway has been considered previously. Kurata and Nagatani [30], using an optimal velocity traffic model, demonstrated the formation of a localized jam just upstream of an accident car (the blockage) and an extended jam at larger incoming flow. Their findings resemble those of Kerner and Klenov [27] for a permanent reduction from two lanes to one lane ("merge bottleneck"). Kerner and Klenov found a transition from free flow to a localized synchronous pattern for incoming flow of about 1000 vehicles/h and another transition at higher flows to an extend phase called the general pattern. Their simulations were based on the three-phase model. Zhu *et al.* [31] used a cellular automata model to study the spatial-temporal profiles, lane change frequencies, and fundamental diagrams for symmetric and asymmetric lane changing rules. They found a jam behind the accident car and a vehicle cluster in the adjacent lane. Yamauchi *et al.* [32] developed a cellular automaton model based on a stochastic optimal velocity model, which they applied to the 2-to-1 bottleneck. They emphasized game theory aspects associated with the decision making of drivers, in particular the advantages and disadvantages of agents using a cooperative strategy compared to agents using a defective strategy (selfish optimization).

The paper is organized as follows. Sec. II contains a description of the Kerner-Klenov model. Results for the travel-time scenario with selfish routing are discussed in Sec. III. Simulations for the hypothetical control scenario are presented in Sec. IV. Decision making based on anticipation is presented in Sec. V and applied to congestion due to boundary conditions in Sec. VI. Conclusions are given in Sec. VII.

**II. Kerner-Klenov model**

In this section, a summary is given of the update, lane changing rules, and diverting to the bypass for the Kerner-Klenov traffic model [24-28] used in simulations presented in this paper.

II.1. *Update rules*

The velocity and position at the next time step ($t+1$) is calculated for any vehicle from its current (at time $t$) velocity $v$ and position $x$ according to the following update rules. The position at $t+1$ is given by

$$x \rightarrow x + v_{new} \qquad (II.1)$$

and the velocity by

$$v \rightarrow v_{new} \qquad (II.2)$$

where

$$v_{new} = \min\{v_{limit}, \tilde{v} + \xi, v + 0.5, v_s\}. \qquad (II.3)$$

The vehicle immediately in front is called the lead vehicle and its velocity and position at $t$ are denoted by $v_l$ and $x_l$. The gap between the vehicles is

$$g = x_l - x - 7.5. \tag{II.4}$$

The function min{} is the minimum of the quantities in brackets. Likewise, max{} denotes the maximum. The speed limit is $v_{\text{limit}}$ and the other quantities are defined below. They are in metric units, *i. e.,* meters, seconds, meter/second, *etc*. Random acceleration and deceleration is given by

$$\begin{aligned} \xi &= 0.5, \ S = 1 \ and \ rnd \leq 0.17, \\ &= -0.5, \ S = -1 \ and \ rnd \leq 0.1, \\ &= 0, \ otherwise. \end{aligned} \tag{II.5}$$

The symbol *rnd* represents a random number in [0,1] and

$$\begin{aligned} S &= -1, \ \tilde{v} < v - 0.01, \\ &= 1, \ \tilde{v} > v + 0.01, \\ &= 0, \ otherwise. \end{aligned} \tag{II.6}$$

where

$$\tilde{v} = \min\{v_s, v_c, v_{limit}\} \tag{II.7}$$

and

$$v_s = \min\{g + v_{la}, v_{safe}\} \tag{II.8}$$

with

$$v_{la} = \min\{g_l, v_l - 0.5, v_{lsafe}\} \tag{II.9}$$

where $g_l$ is gap between the lead vehicle and the one in front of it and $v_{lsafe}$ is the "safe" velocity of the lead vehicle analogous to $v_{safe}$ defined below, Eqs. (II.10-16). It is introduced to avoid collisions.

$$v_{safe} = \alpha_{safe} + \beta_{safe} \tag{II.10}$$

and

$$\alpha_{safe} = Int(z) \tag{II.11}$$

where $Int(z)$ denotes the integer part of $z$ and

$$\beta_{safe} = \frac{X}{\alpha_{safe} + 1} - 0.5\alpha_{safe}, \tag{II.12}$$

where

$$X = \alpha\beta + 0.5\alpha(\alpha - 1) + g, \tag{II.13}$$

$$\alpha = Int(v_l), \tag{II.14}$$

$$\beta = v_l - \alpha, \tag{II.15}$$

and

$$z = \sqrt{2X + 0.25} - 0.5. \tag{II.16}$$

Furthermore

$$v_c = v + a \quad if \quad g > G,$$

$$= v + \Delta, \quad otherwise, \tag{II.17}$$

where the synchronization length is

$$G = 3v + 2v(v - v_l) \tag{II.18}$$

and the stochastic time delay of acceleration and deceleration is represented by

$$\Delta = \max\{-b, \min\{v_l - v, a\}\}. \tag{II.19}$$

To calculate *a* and *b,* let

$$p_0 = 0.575 + 0.125 \min\{\frac{v}{10}, 1\}, \tag{II.20}$$

$$p_2 = 0.48, \quad v \leq 15,$$

$$= 0.8, \quad v > 15, \tag{II.21}$$

$$P_0 = 1, \quad S^{prev} = 1,$$

$$= p_0, \quad otherwise, \tag{II.22}$$

and

$$P_1 = p_2, \quad S^{prev} = 1,$$

$$= 0.3, \quad otherwise. \tag{II.23}$$

$S^{prev}$ is the quantity $S$, Eq. (II.6), calculated for this vehicle in the previous time step. Then

$$a = 0.5, \quad rnd < P_0,$$

$$= 0, \quad otherwise, \tag{II.24}$$

and

$$b = 0.5, \quad rnd < P_1,$$

$$= 0, \quad otherwise. \tag{II.25}$$

If a vehicle is in the exit region of the right lane near an off-ramp, $v_c$ (but not $v_s$) is calculated using the position of the vehicle immediately ahead of it on the bypass, $x_{ahead}$, if one exists. The velocity of the lead vehicle in the calculation of $v_c$ is taken to be $v_{ahead} - 2$ or the bypass speed limit, if lower.

The following variables are required to be non-negative: $v_{new}$, $\tilde{v}$, $v_{la}$, $\alpha_{safe}$, and $G$.

II.2. *Rules for lane changing*

Since the main line consists of two lanes, the rules for changing lanes are given below. The vehicles forming the gap in the target lane are denoted by "ahead" and "follow" while the vehicle immediately in front in the same lane is the "lead" vehicle at $x_l$ with velocity $v_l$. The position and velocity of the vehicle considered for a lane change are $x$ and $v$.

The motivation conditions are:

$$v \geq v_l \tag{II.26}$$

and

$$v_{ahead} \geq v_l + 2. \tag{II.27}$$

Then let

$$g^+ = x_{ahead} - x - 7.5, \tag{II.28}$$

$$g^- = x - x_{follow} - 7.5,  \quad (\text{II}.29)$$

$$G^+ = \min\{G_{ahead}, v\}, \quad (\text{II}.30)$$

and

$$G^- = \min\{G_{follow}, v_{follow}\}, \quad (\text{II}.31)$$

where

$$G_{ahead} = 3v + 2v(v - v_{ahead}) \quad (\text{II}.32)$$

and

$$G_{follow} = 3v_{follow} + 2v_{follow}(v_{follow} - v). \quad (\text{II}.33)$$

The security conditions require

$$g^+ > G^+ \quad (\text{II}.34)$$

and

$$g^- > G^-. \quad (\text{II}.35)$$

If all conditions are satisfied, a lane change occurs if

$$rnd \leq p_c = 0.45.$$
(II.36)

Note that this is an earlier version of $p_c$. Later papers [24, 26] used $p_c = 0.2$. In the calculations presented in this paper, if the gap ahead in the target lane or the same lane is larger than 150 m, then the speed of the leading vehicle is taken as effectively infinite.

II.3. *Taking the bypass*

If drivers decide to take the bypass (or in a control situation are directed to exit) and are in the off-ramp region of the right lane, they do so if the space ahead is large enough. The required distance between the last vehicle on the bypass and the vehicle about to take the bypass is its velocity $v$ times a headway time $h_d = 1$ s plus the length of a vehicle, $\ell = 7.5$ m. That is,

$$x_{last} - x > v h_d + \ell$$
(II.37)

More than one vehicle can exit at the same time.

**III. Travel-time scenario**

In this section, drivers choose one of two routes based on travel-time information. This is the selfish routing case. The two routes are shown in Fig. 1. The first route, the main line, is a two-lane freeway with a stalled vehicle in the right lane. The second route, called the bypass, is a lower-speed road such as a surface street connecting the same origin and destination; these are a distance $L$ apart. The stalled vehicle is at a distance $X_{stall}$ from the origin. The length of the off-ramp to the bypass is taken to be 500 m. The speed limits on the routes are 32 m/s and 25 m/s, unless otherwise noted.

The time to travel from the origin to the destination is denoted $T_{path}^{main}$ for the main line and $T_{path}^{bypass}$ for the bypass. To make these travel times somewhat smooth, they are averaged in the following way.

$$T_{path} = \alpha T + (1-\alpha)T_{path}^{prev}. \qquad (III.1)$$

For either route, the smoothed travel time is updated every time a vehicle completes that route with travel time $T$. The value chosen for $\alpha$ is a compromise between reasonable smoothing so that drivers are not confused by wild fluctuations and the requirement to be as current as possible. It was found that $\alpha = 0.1$ is a suitable choice.

In Fig. 2, the incoming flow to $x = 0$, the origin, is shown. The flow gradually increases from 425 vehicles/h/lane to 850 vehicles/h/lane over about one-half hour, after which time the flow is constant. This input flow will be used throughout this section.

Results are shown in Fig. 3 for $X_{stall} = 2$ Km and $L = 3$ Km. The top curve in Fig. 3a is for the main line. Characteristic oscillations similar to those observed by Wahle *et al.* [9,10] can be seen. The next curve, representing the travel time on the bypass, is essentially flat. Because the bypass is uniform and flow rates are never large, the travel time on the bypass is always quite close to $L/25$ m/s. The next two curves are the number vehicles on the main line and the bypass respectively. Like $T_{path}^{main}$, these too show characteristic oscillations.

The close-up of $T_{path}^{main}$ and $N_{main}$ for 6000 s $< t <$ 7500 s (Fig. 3b) indicates the rise in $N_{main}$ precedes the increase in $T_{path}^{main}$ by an amount of the order of the delay, *i.e.*, the travel time. The period of the oscillations is much larger (~1000 s) and is primarily determined by the time needed to clear the congestion near the stalled car and to restore free flow on the main line.

A number of examples with different $X_{stall}$, $L$, and incoming flow have been investigated. If main-line congestion causes drivers to take the bypass, characteristic oscillations always develop when there is selfish routing based on travel-time information. It appears to be a general feature. Without some kind of learning or anticipation, selfish routing does not appear to approach a Wardrop equilibrium with equal travel times on the two routes.

## IV. Hypothetical control scenario

Because it was found that the number of vehicles on the main line $N_{main}$ leads (in time) $T_{path}^{main}$, simulations in Sec. III suggest that controlling $N_{main}$ and $N_{bypass}$ might be useful in obtaining an equilibrium. Also, $N_{main}$ and $N_{bypass}$ can be changed almost immediately by either directing vehicles to exit to the bypass or remain on the main line. In this paper, no attempt to describe a practical system to accomplish such control is presented. It is assumed, for the purpose of evaluating the potential to establish a Wardrop equilibrium, that such control is possible and that drivers respond faithfully to control signals. Thus, in this section, vehicles in the right lane exit to the bypass if they are in the off-ramp region, satisfy the requirement (II.37), and

$$N_{bypass} < N_{main}. \qquad (IV.1)$$

Exiting to the bypass initially commences when $T_{path}^{main}$ first exceeds $T_{path}^{bypass}$.

As demonstrated in Fig. 4, this control scheme establishes steady flow with $T_{path}^{main}$ lower than $T_{path}^{bypass}$. Results for three different values of $X_{stall}$ (2, 4, and 6 Km with $L = X_{stall} + 1$ Km) are displayed for the incoming flow of Fig. 2. The dashed lines are the nearly constant travel times for the bypass and the solid curves are for the main line. Although a Wardrop equilibrium is not obtained, the average travel time for all vehicles is less for the steady state obtained than it would be if a Wardrop equilibrium were realized.

It is straightforward to show that when $N_{bypass} = N_{main}$ the average travel time for all vehicles is

$$\overline{T}_{path} = \left[0.5\left(1/T_{path}^{main} + 1/T_{path}^{bypass}\right)\right]^{-1}_{control} < \overline{T}_{path}^{bypass}\bigg|_{Wardrop}. \tag{IV.2}$$

The inequality follows because, for there to a Wardrop equilibrium, the main-line travel time would have to increase to that of the bypass, which remains approximately $L/25$ m/s.

The steady state obtained when $N_{bypass} = N_{main}$ is not necessarily a system optimum, but it appears to be rather close. To reduce $\overline{T}_{path}$ further would require diverting fewer vehicles to the bypass without significantly increasing $T_{path}^{main}$.

In an attempt to do so, requiring vehicles to take the bypass is changed from the condition (IV.1) to whenever

$$N_{bypass} < N_{main} - N_{off} \tag{IV.3}$$

where $N_{off}$ is an offset. Results for the case of $X_{stall} = 2$ Km are shown in Fig. 5 with $N_{off} = 0, 12, 15,$ and 18. In each panel, the blue curve is $T_{path}^{main}$. The nearly flat curve is $T_{path}^{bypass}$ and the red and light blue curves are $N_{main}$ and $N_{bypass}$, respectively. For large $N_{off}$, $T_{path}^{main}$ exceeds $T_{path}^{bypass}$ after a half hour. For $N_{off} = 12$ or 15, $T_{path}^{main}$ oscillates irregularly around $T_{path}^{bypass}$, but no equilibrium is obtained.

The average travel time as a function of $N_{off}$ for vehicles on the main line during the last hour of the simulation is shown in Fig. 6. Average $T_{path}^{main}$ equals that of the bypass for $N_{off}$ in the range 12 to 15. A broad minimum in the average travel time for all vehicles (not shown) occurs near $N_{off} \approx 0$. Simulations for negative $N_{off}$ show no further decrease in $T_{path}^{main}$.

The reason the control law (IV.1) works is because enough incoming flow is diverted to the bypass to avoid a transition from free flow to a congested state near the stalled vehicle. The transition is demonstrated in Figs. 7 and 8. For these calculations there is no bypass. The constant incoming flow is equal in each lane and encounters the stalled vehicle at 2 Km from the origin. The travel time from the origin to 3 Km is calculated for various incoming flows (all on the main line). The average over the second hour of simulation is shown as a function of incoming flow in Fig. 7. An abrupt increase in the average travel time occurs between 455 and 460 vehicles/h/lane. Curves of $T_{path}$ as a function of time for the two flows are shown in Fig. 8. The upper curve corresponds to the congested phase and the lower to the free-flow phase that the system fluctuates between in the middle two panels of Fig. 5 for $N_{off}$ = 12 to 15.

A Wardrop equilibrium can not be obtained because a phase transition exists. For there to be equal travel times, the main-line flow would have to be intermediate between the congested and free-flow phases, which appears to be unstable. Similar transitions from free to congested flow have been found previously for other traffic flow models: the optimal velocity model [30] and a cellular automata model [31]. Thus it is expected that phenomena reported in the present work would be observed in simulations with other traffic models.

For incoming flow low enough to avoid forming a congested phase (< 455/h/lane), diverting to the bypass is unnecessary. For sufficiently large incoming flow (*e.g.*, 1000/h/lane), diversion to the bypass cannot prevent a congested state because the bypass removes at most the incoming flow in the right lane. No special provision (beyond ordinary lane changing) for vehicles to go into the right lane in order to take the bypass is included in the model.

The results of this section can be related to those for selfish routing discussed in Sec. III. In Fig. 3, the system also appears to fluctuate between two phases. Because diverting to the bypass is delayed, the number of vehicles on the main line builds up to a larger value than seen in Fig. 5 and consequently $T_{path}^{main}$ reaches larger values. It is interesting to note that for a half-hour interval near $t$ = 4000 s $N_{main}$ and $N_{bypass}$ are nearly equal. This is not the system optimal state, however, because $T_{path}^{main}$ is greater than $T_{path}^{bypass}$. It seems that the main line remains in the congested phase described above (Figs. 5 and 8) for an extended time before finally dissipating.

Because vehicles that divert to the bypass can in practice take several possible routes to the destination and other vehicles can enter the bypass from side streets, $N_{bypass}$ might not be a viable or convenient quantity to compare to $N_{main}$. An alternative control method is to control $N_{main}$ to a target value, $N_{t\arg et}$. In Fig. 9, an example is shown for a constant incoming flow of 750/h/lane with $X_{stall}$ = 2 Km and $L$ = 3 Km. With the target set at $N_{t\arg et}$ = 25 vehicles, the congested state dissipates after the initial rise of $T_{path}^{main}$ above $T_{path}^{bypass}$. The travel time on the main line $T_{path}^{main}$ drops below $T_{path}^{bypass}$ and mostly remains lower. However, if the target is set slightly higher, $N_{t\arg et}$ = 30 vehicles, the congested phase persists and $T_{path}^{main}$ oscillates somewhat irregularly about $T_{path}^{bypass}$ (See Fig. 10.). The cumulative average travel time $T_{ave}$ for all vehicles reaching the destination is shown in Fig. 11 for $N_{t\arg et}$ = 25 and 30 vehicles. Clearly the average travel time for the lower target value is closer to the system optimal value.

For $N_{t\arg et}$ = 30 vehicles, the probability that taking the bypass results in a shorter travel time compared to remaining on the main line is fairly high. In Fig. 12, the probability $P$ of actual travel times for vehicles on the main line and the bypass are shown. Although some main-line travel times are shorter, many are higher than those for the bypass. The average main-line time is 120 s whereas the bypass time is 119 s. Thus, drivers required to take the bypass do not feel penalized and the routing is fair. In contrast, when $N_{t\arg et}$ = 25 vehicles, the probability of travel times shown in Fig. 13 indicates that taking the bypass often results in a longer travel time; the average main-line time is now 111 s and the bypass remains at 119 s. Drivers could justifiably feel the routing is unfair, even though it is close to a system optimum. Thus we have the classic dilemma of the unfairness of the system optimal solution relative to an equilibrium where all drivers experience the same travel times. Admittedly the differences are small in this case and might not make a significant practical difference, but these examples illustrate the dilemma. For longer paths or if the bypass is a lower-speed route (<< 25 m/s), fairness might be a consideration.

**V. Anticipation**

In this section, anticipation is used to advise drivers when to divert to the bypass. This involves collecting data in the following manner. When a vehicle $n_i$ leaves the origin, record the number of vehicles on the main line $N_{main}^{start}(n_i)$ at that time; and when the vehicle reaches the destination, record the actual travel time $T_{main}^{travel}(n_i)$ if the vehicle remained on the main line. [Recall that $T_{path}^{main}$ is a smoothed travel-time function, see Eq. (III.1). It is to be distinguished from $T_{main}^{travel}(n_i)$.] Compare the travel time to the bypass travel time, which is taken for simplicity to be $T_{bypass} = L/v_{lim}$ where $v_{lim}$ is the speed limit on the bypass (generally 25 m/s). Sort the data into two groups according to whether $T_{main}^{travel}(n_i) > T_{bypass}$ or not.

At any time $t$ after $T_{path}^{main}$ first exceeds $T_{bypass}$ (the first crossing), the average of the main-line number of vehicles that resulted in a travel time on the main line greater than the bypass is calculated:

$$N_> = \frac{\sum_{i=1}^{M_t} N_{main}^{start}(n_i)}{M_t} \qquad (V.1)$$

where $M_t$ is the number of vehicles reaching the destination for which $T_{main}^{travel}(n_i) > T_{bypass}$ by time $t$. The average is a function of time but tends to a nearly constant value as time goes on.

In addition to computing the average $N_>$, the complimentary average $N_<$ is defined as

$$N_< = \frac{\sum_{i=1}^{M_t'} N_{main}^{start}(n_i')}{M_t'} \qquad (V.2)$$

where $M_t'$ is the number of vehicles at time $t$ for which $N_{main}^{start}(n_i')$ leads to a travel time $T_{main}^{travel}(n_i') \leq T_{bypass}$.

The averages $N_>$ and $N_<$ are, given the information available, the best indicators of whether or not a driver should take the bypass. After the first crossing, $T_{path}^{main}$ is less reliable (leads to large oscillations shown in Fig. 3) and is not used any further. Thus, for vehicle $n$ at the origin, if

$$N_{main}^{start}(n) > N_> \qquad (V.3)$$

the driver anticipates that remaining on the main line would likely result in a longer travel time than taking the bypass. Likewise, if

$$N_{main}^{start}(n) < N_< \qquad (V.4)$$

the driver of vehicle $n$ would not take the bypass. If $N_{main}^{start}(n)$ is in between $N_<$ and $N_>$, the driver must decide which of two roughly equivalent paths (as regards expected travel time) to take. In the following analyses, a simple assumption that the driver chooses to take the bypass only if

$$N_{main}^{start}(n) > \frac{1}{2}(N_< + N_>) \qquad (V.5)$$

is made.

In Fig. 14, results are shown for the incoming flow depicted in Fig. 2 and the parameters pertaining to Fig. 3. The main-line travel time $T_{path}^{main}$ is shown in blue. Once drivers begin to use anticipation for deciding to divert to the bypass ($t > 441$ s), $T_{path}^{main}$ soon falls below $T_{bypass}$ (at $t > 652$ s) and mostly remains lower. The average value of main-line travel times over two hours is 109 s, about 10% smaller than $T_{bypass} = 120$ s. The number of main-line vehicles initially rises, levels off, and then approaches ½ ($N_< + N_>$) ≈ 23 vehicles and fluctuates closely about this number thereafter. The value that ½ ($N_< + N_>$) approaches is less than the target value of 25 that prevented a transition to a congested state in Fig. 9.

The results are similar for larger values of $X_{stall}$ and $L$ (6 Km and 8 Km, respectively) shown in Fig. 15. After about ½ h, $T_{path}^{main}$ (blue) drops below $T_{bypass}$ (red) and fluctuates between 265 and 320 s. The cumulative average travel time (green) was 287 s after two hours. The number of vehicles on the main line $N_{main}$ is shown as a function of time in Fig. 16. After 1500 s, $N_<$ and $N_>$ gradually approach one another and nicely bracket $N_{main}$.

Although the average travel time for drivers taking the bypass was larger than the average main-line travel time in these examples, the system can still be considered fair because drivers chose on the basis of the best information available at the time they encountered the bypass. The system reached a steady state that was not a Wardrop equilibrium nor a system optimum, but something in between. If one compares Fig. 14 to Fig. 3 (for $L = 3$

Km) or Fig. 17 to Fig. 15 ($L = 8$ Km), it can be seen that decision making based on anticipation (Figs. 14 and 15) using the number of main-line vehicles is preferable to using current values of $T_{path}^{main}$ relative to $T_{path}^{bypass}$ (Figs. 3 and 17).

Similar results are found for $X_{stall} = 2$ Km and $L = 3$ Km, but with a lower speed limit on the bypass, $v_{lim} = 20$ m/s. The results are displayed is Fig. 18. Here $T_{ave}$ for two hours is 130 s compared to 150 s for the bypass. (In Fig. 14 with $v_{lim} = 25$ m/s, $T_{ave} = 109$ s reflecting the lower bypass travel time of 120 s.) However, the rate of diverting to the bypass, shown in Fig. 19, is high only when $T_{path}^{main}$ is close to $T_{bypass}$ (2000 to 4000 s) and decreases significantly during intervals where $T_{path}^{main}$ is lower ($t > 4000$ s). Thus, the majority of vehicles taking the bypass are not significantly penalized relative to remaining on the main line and the system is fair.

A final example is for higher flow with a rapid increase from 500/h/lane to 1000/h/lane (over just 100) cars followed by steady incoming flow. Because the flow is so large, it is not possible to divert enough of the flow to prevent a transition to the congested phase. The best that can be done is to divert all those vehicles incoming on the right lane. Fig. 20 shows the results for $X_{stall} = 2$ Km and $L = 3$ Km. During the last ½ h, vehicles diverted at a rate of 984/h; nearly all those in the right lane took the bypass as needed to obtain the best results. Those drivers taking the bypass were not penalized because they had shorter travel times than those remaining on the main line (120 s compared to 157 s).

**VI. Congestion due to boundary conditions**

Rather than considering congestion due to a stalled vehicle at $X_{stall} < L$, in this section a low-speed boundary condition is imposed on the main line at the destination. Specifically, vehicles are required to slow to the speed $u_{BC}$ which is less than the speed limit. No boundary condition is applied to the bypass. The reason for considering such congestion is that route travel time is more nearly a continuous function of the number of vehicles on the path than a route with a stalled vehicle. There is no abrupt jump due to a phase transition. Conditions are more like those typically analyzed when finding Wardrop equilibria.

In Fig. 21 vehicle velocity is plotted against vehicle position for $L$ = 8 Km. The incoming flow was 2400/h/lane (after gently increasing from 1200/h/lane over 500 cars). With $u_{BC}$ = 20 m/s, a region of synchronous flow developed for ~ 6 Km upstream of the boundary. Also, a wave of reduced speed originating from vehicles slowing to take the bypass can be seen for – 8 Km < $x$ < - 4 Km. The snapshot of velocity *vs* position is for $t$ = 2 h in a simulation where drivers use anticipation to decide which route, main line or bypass, to take.

Travel time as a function of time is shown in Fig. 22. At about ½ hour $T_{path}^{main}$ exceeded $L$/ 25 m/s = 320 s and drivers began to take the bypass. The travel time then dropped somewhat below 320 s and gradually approached the bypass time. The averages $N_<$ and $N_>$ are displayed in Fig. 23 along with $N_{main}$, which closely followed ½ ($N_<$ + $N_>$). The resulting steady state is nearly a Wardrop equilibrium with equal route travel times and constant vehicle number, which implies constant flow at a rate of $N_{main} / T_{path}^{main}$. The corresponding plot of $T_{path}^{main}$ against time for selfish routing based on current travel times is shown in Fig. 24. The characteristic oscillations seen in previous scenarios also occur in this example.

If the boundary speed is lowered to $u_{BC}$ = 10 m/s, $T_{path}^{main}$ becomes virtually identical to the bypass time of 320 s in less than one hour. Other simulations with different parameters (not shown as well) indicate that anticipation produces steady states closely resembling the putative Wardrop equilbria.

## VII. Conclusions

In this paper, simulations were done to examine how real-time information on travel times for alternative routes might establish a Wardrop equilibrium. A realistic, stochastic model of traffic flow was used to examine congestion caused by a stalled vehicle and the effect of drivers diverting to a lower-speed bypass. The traffic model was the three-phase model of Kerner and co-workers [24-28], but the results are thought to be more general because other models give a similar transition to a congested phase on the main line.

The first scenario considered was that of selfish routing in which each driver chose a route to take (a two-lane main line with a stalled vehicle or a bypass) based upon current values of the travel time. Strong oscillations in main-line travel time occurred in a manner like that discussed by Wahle *et al.* [9,10] The period of oscillation was found to be related to the time required to clear congestion when main-line flow deceased as vehicles diverted to the bypass. The reason oscillations occurred was attributed (as in Wahle *et al.*) to the inherent delay in the information, which could only be as recent as the last vehicle to complete the route. No equilibrium was found with selfish routing.

Following the observation that travel time lags the number of vehicles on the main line, hypothetical control methods based on the number of vehicles on the routes were proposed. In these methods, the number of main-line vehicles was controlled to either the number on the bypass minus an offset or to a predetermined target value. A steady state in which a transition to the congested state was avoided could be obtained with either method using the proper value of the offset or target. The main-line travel time established was generally less than the nearly constant bypass time, although the latter could be adjusted by changing the speed limit to coincide with the main-line travel time. A steady state that was close to the system optimum could be obtained, but not a Wardrop equilibrium.

If sufficient flow were diverted to the bypass, main-line travel time was nearly constant. However, if the incoming flow were large enough (~ 1000 vehicles/h/lane), congestion set in and travel time abruptly increased. Roughly speaking, two values of travel time were found for a given route length: one for free flow and another for the congested phase. If the offset or target values were taken to be on the boundary between those corresponding to the two phases, travel time was found to fluctuate irregularly between the free-flow and congested-flow characteristic times. Over an interval of several hours, the average travel time on the main line might coincidently equal that of the bypass, but no true Wardrop equilibrium was found. Control measures that avoided the free flow-congested phase transition, although approximating a system optimum, could be considered unfair because drivers were required to divert at times when the bypass travel time exceeded that of the main line.

A third scenario involving anticipation was analyzed to find a driver-advice system that could be considered fair, yet would avoid the deficiencies of selfish routing or control. The essential concept of the proposed system was that the number of main-line vehicles on the route at the time a driver leaves the origin is the pertinent variable in predicting travel time. Compared to using current travel time, drivers responding to the anticipation of travel time produced more acceptable results. Due to considering incoming flow near the phase transition between free and congested states, the anticipation scheme did as well as could be expected, yet demonstrated the difficulty in attaining a steady state resembling a Wardrop equilibrium with a stalled vehicle. For flow not involving a phase transition where travel time is essentially a continuous function of the number of main-line vehicles (such as replacing the stalled vehicle with a low-speed boundary condition), anticipation worked remarkably well and produced a steady state virtually identical to the Wardrop equilibrium.

**Figures**

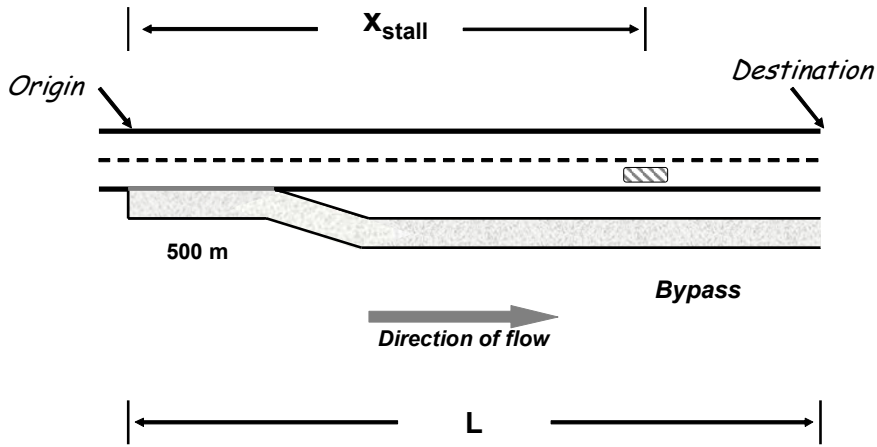

Fig. 1. Two routes connecting the origin and the destination, a distance *L* apart. The main line consists of two lanes with a stalled vehicle a distance $X_{stall}$ from the origin. The off-ramp to the bypass is 500 m in length. The speed limits are 32 m/s and 25 m/s, unless otherwise noted.

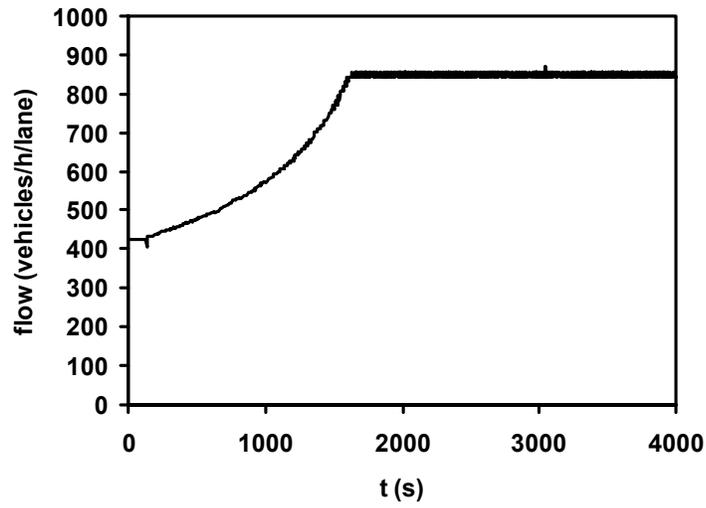

Fig. 2. Typical incoming flow as a function of time. The flow increases from 425 vehicles/h/lane to 850 vehicles/h/lane over 500 cars (250 in each lane) in ½ h.

(a)

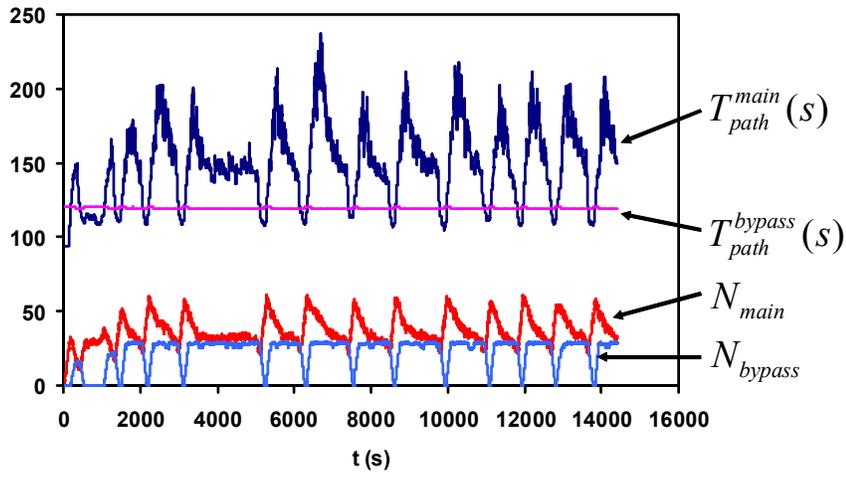

(b)

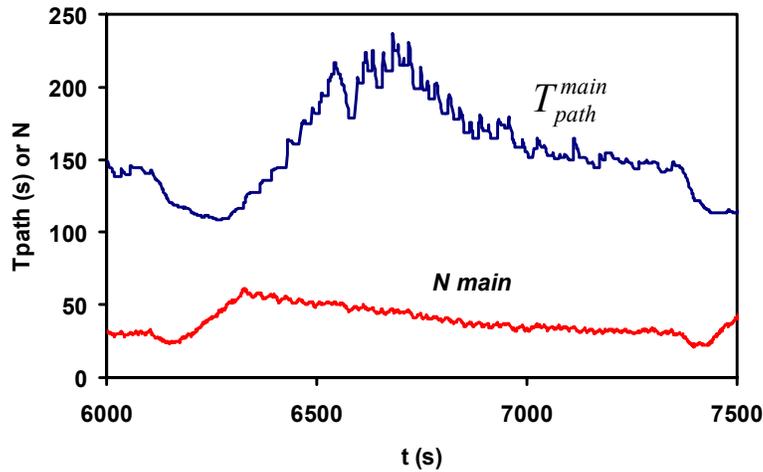

Fig. 3. Travel time and number of vehicles on each route for incoming flow shown in Fig. 2 with $X_{stall}$ = 2 Km and $L$ = 3 Km and selfish routing based on comparing $T_{path}^{main}$ to $T_{path}^{bypass}$ [See Eq. (III.1) for definitions.]. (a) Results for both routes over four hours. (b) Travel time and number of vehicles on the main line for 6000 s < $t$ < 7500 s.

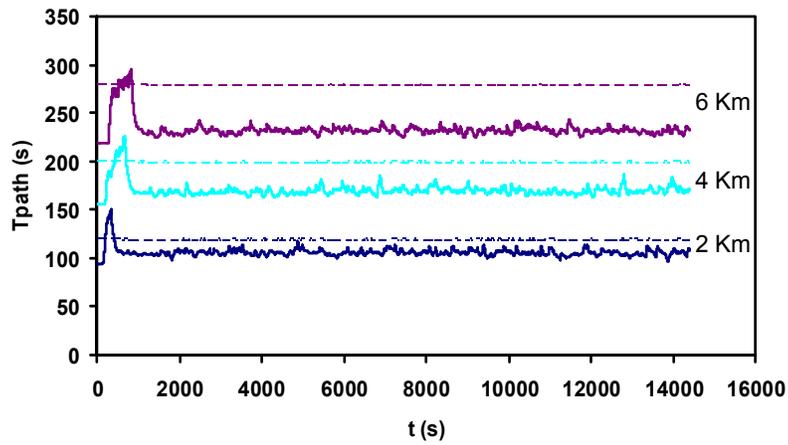

Fig. 4. Travel time for main line (solid curves) and bypass (dashed lines) for $X_{stall}=$ 2, 4 and 6 Km and $L = X_{stall} + 1$ Km. The incoming flow is shown in Fig. 2. Vehicles in the right lane are required to take the bypass by a hypothetical control system that closely maintains the same number of vehicles on each route. [See Eq. (IV.1) and related text.]

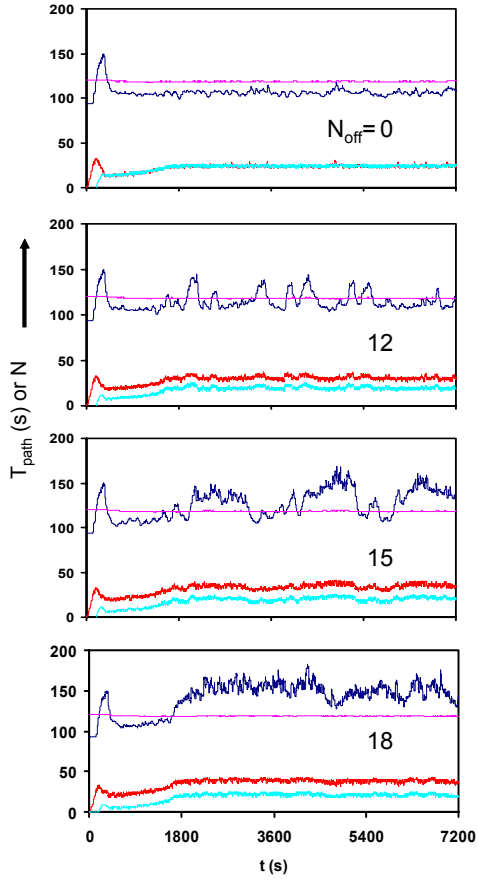

Fig. 5. Travel time and number of vehicles on each route for incoming flow shown in Fig. 2, $X_{stall}$ = 2 Km and $L$ = 3 Km and control such that $N_{main} = N_{bypass} + N_{off}$. In each panel, the blue curve is $T_{path}^{main}$ and the nearly flat curve is $T_{path}^{bypass}$. At the bottom of each panel, the red and light blue curves are $N_{main}$ and $N_{bypass}$, respectively. [ See Eq. (IV.3) and related text.]

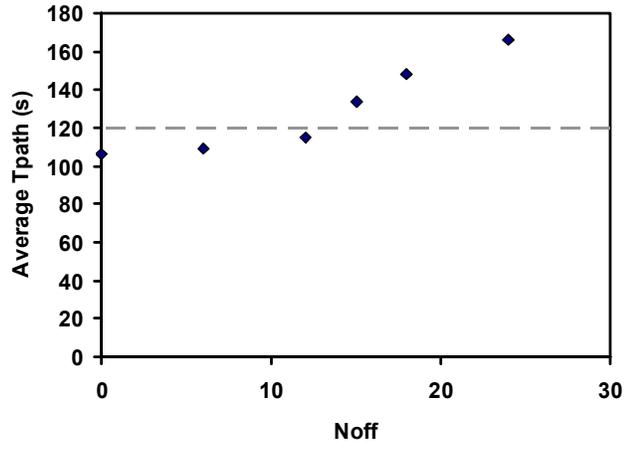

Fig. 6. Average main-line travel time during last hour of Fig. 5 *vs* $N_{off}$. The dashed line is the bypass travel time.

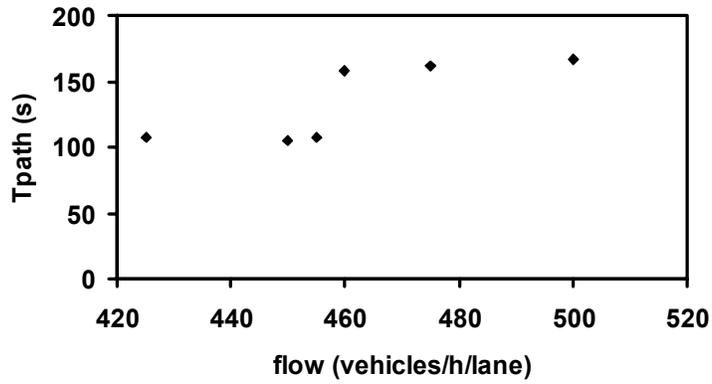

Fig. 7. Average main-line travel time (over 1 h < $t$ < 2 h) *vs* incoming flow with no diversion to bypass. The incoming flow is constant. Note the abrupt increase between 455 and 460 vehicles/h/lane. $X_{stall}$ = 2 Km and $L$ = 3 Km.

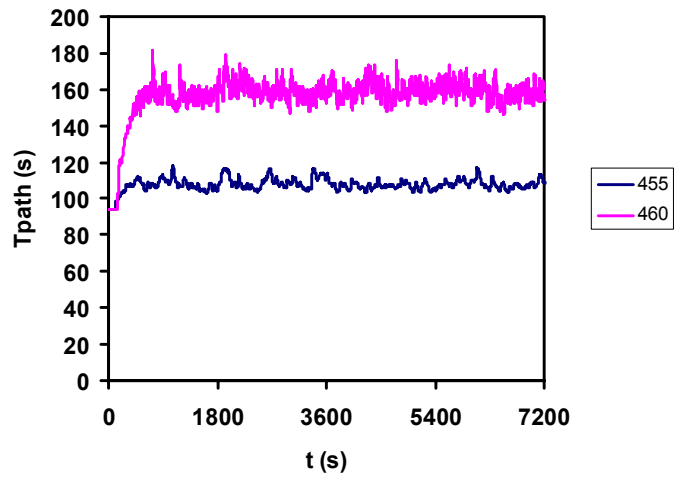

Fig. 8. Main-line travel time as a function of time for two incoming flows and no diversion to the bypass. The blue curve is for 455 vehicles/h/lane and the red for 460 vehicles/h/lane corresponding to the abrupt increase in $T_{path}^{main}$ noted in Fig. 7. A phase transition from free flow to a congested state upstream of the stalled vehicle occurs. $X_{stall}$ = 2 Km and $L$ = 3 Km.

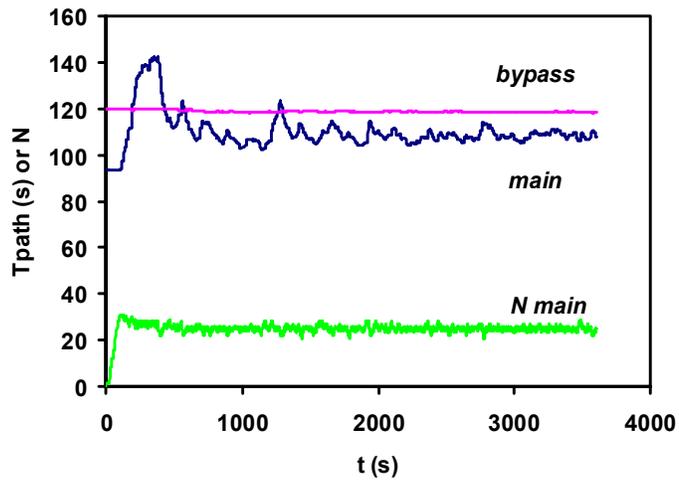

Fig. 9. Travel time and number of vehicles on the main line as a function of time for constant incoming flow of 750 vehicles/h/lane, $X_{stall}$ = 2 Km and $L$ = 3 Km, and a hypothetical control system to maintain $N_{main} = N_{target}$. The target value is set to $N_{target} = 25$.

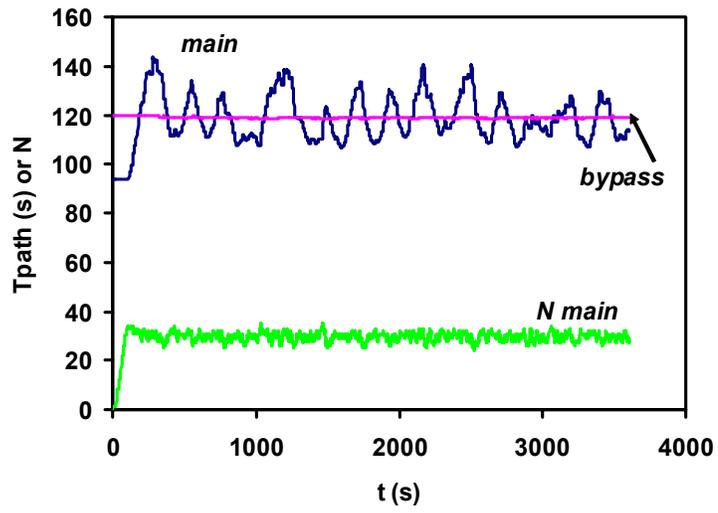

Fig. 10. Same as Fig. 9, but with $N_{target} = 30$. The larger target causes $T_{path}^{main}$ to fluctuate between a free-flow and a congested-flow characteristic travel time.

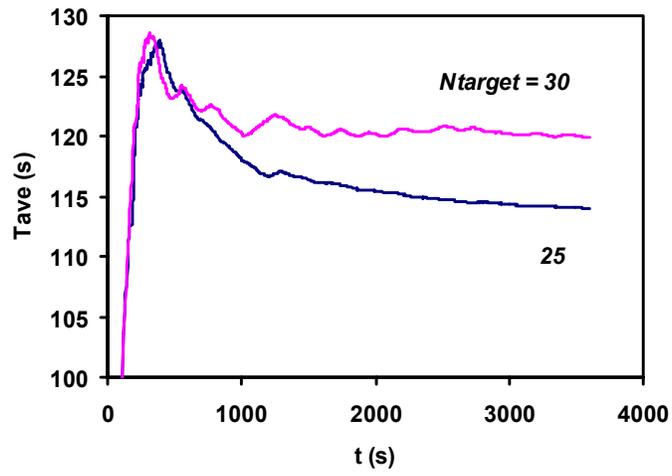

Fig. 11. The cumulative average travel time for the main line for $N_{target} = 25$ and 30. See Figs. 9 and 10.

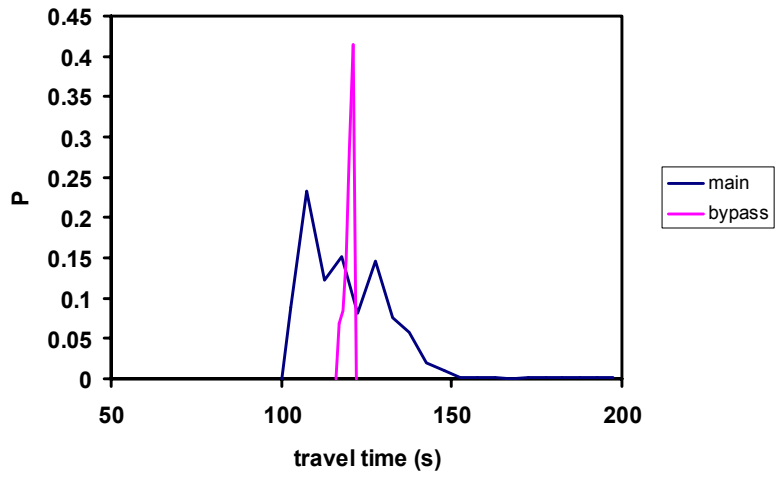

Fig. 12. The distribution of actual travel times on the main line and the bypass for $N_{t\arg et} = 30$. See Fig. 10.

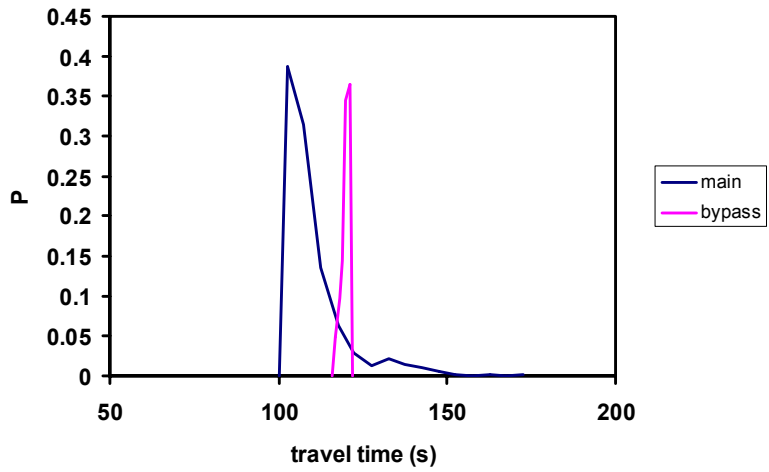

Fig. 13. The distribution of actual travel times on the main line and the bypass for $N_{target} = 25$. See Fig. 9.

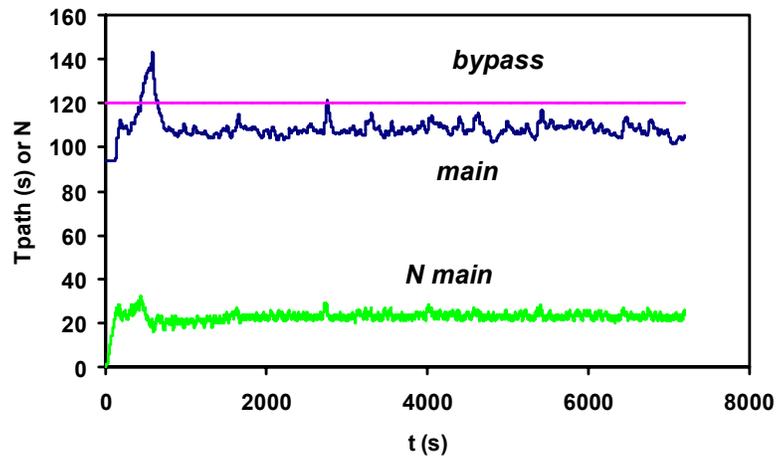

Fig. 14. Travel time and number of vehicles on the main line for drivers using anticipation to determine the shorter route time. The incoming flow is shown in Fig. 2 and $X_{stall}$ = 2 Km and $L$ = 3 Km.

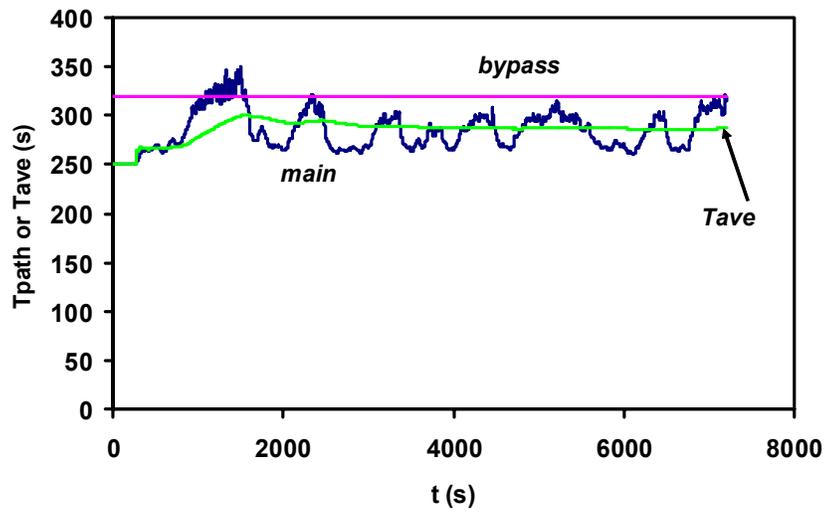

Fig. 15. Travel time for drivers using anticipation to determine shorter route times. The incoming flow is shown in Fig. 2 and $X_{stall}$ = 6 Km and $L$ = 8 Km. The blue curve is for the main line, the red for the bypass, and the green is the cumulative average travel time on the main line.

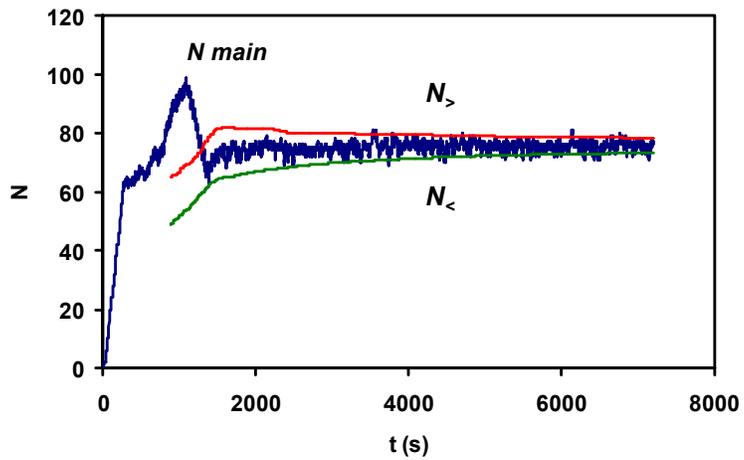

Fig. 16. The number of vehicles on the main line (blue curve) for the parameters of Fig. 15. The red and green curves are the averages $N_>$ and $N_<$, respectively. [See Eqs. (V.1) and (V.2) for definitions.]

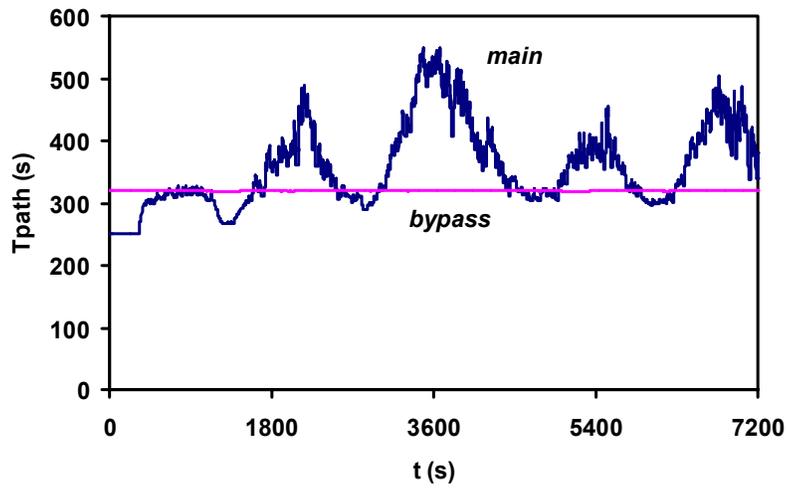

Fig. 17. The travel time as a function of time for the main line and bypass for the parameters of Fig. 15. Drivers choose routes on the basis of selfish routing comparing $T_{path}^{main}$ to $T_{path}^{bypass}$.

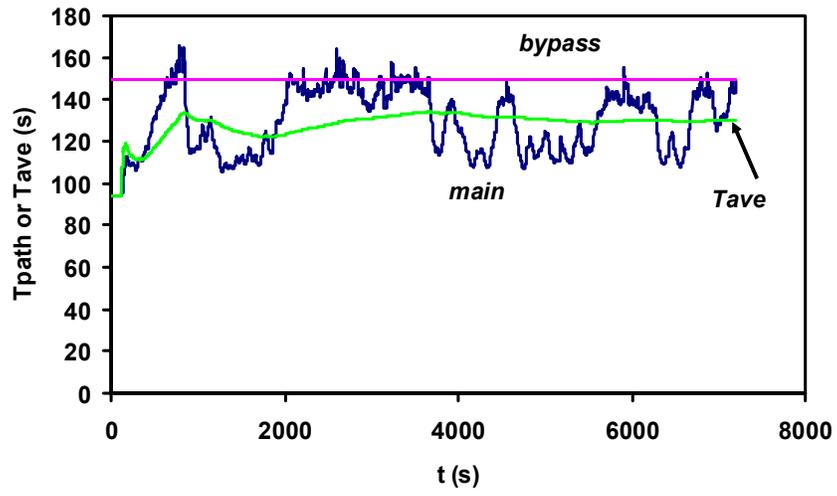

Fig. 18. Same as Fig. 15, but with the bypass speed limit reduced to 20 m/s (rather than 25 m/s).

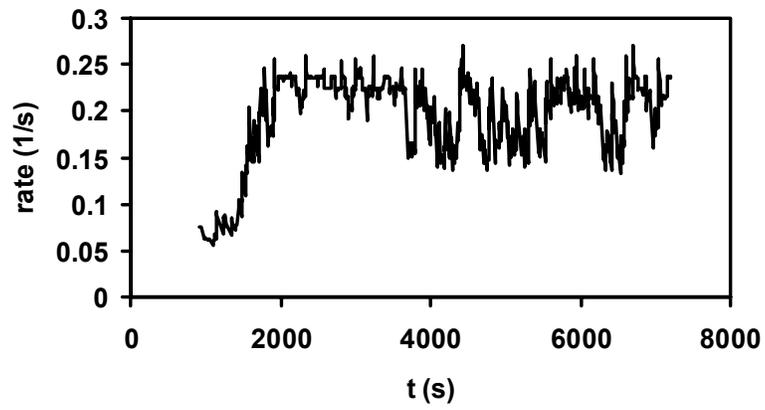

Fig. 19. The rate of taking the bypass as a function of time for the simulation of Fig. 18. The rate is a moving twenty-car average.

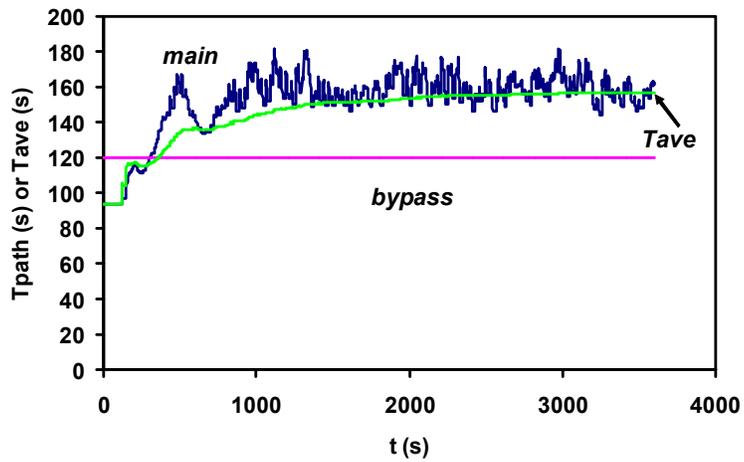

Fig. 20. Travel time for drivers using anticipation to determine shorter route times. The incoming flow increases from 500 vehicles/h/lane to a final value of 1000 vehicles/h/lane over 100 vehicles in each lane; $X_{stall}$ = 2 Km and $L$ = 3 Km. The blue curve is for the main line, the red for the bypass, and the green is the cumulative average travel time for the main line. Since 98% of vehicles in the right lane took the bypass during the last ½ h, the main line travel time could not be lowered to match the bypass.

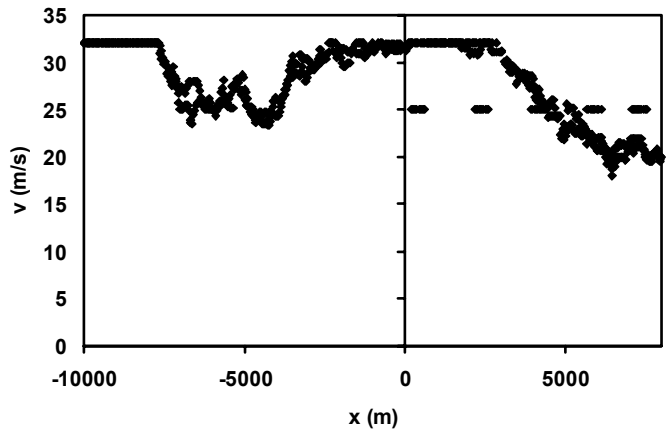

Fig. 21. Velocity of vehicles as a function of their position for a simulation with no stalled car. Instead a boundary condition is imposed at *L*, where vehicles are required to reduce their speed to 20 m/s on the main line. The incoming flow increased from 1200 to 2400 vehicles/h/lane over 1000 cars (500 in each lane). Drivers use anticipation to choose their route.

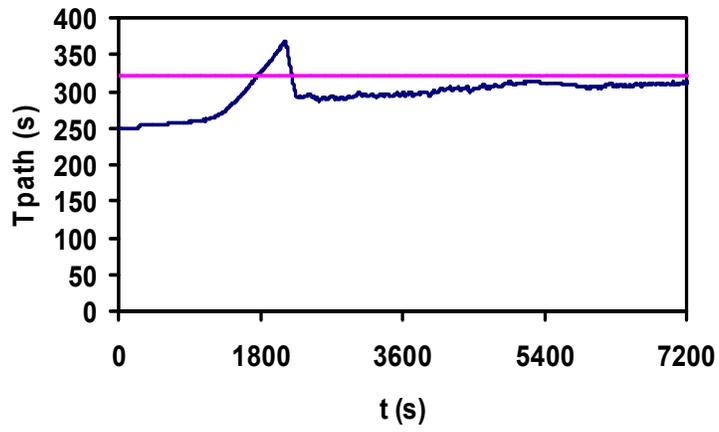

Fig. 22. The travel time (blue curve) $T_{path}^{main}$ as a function of time for the simulation of Fig. 21. The bypass estimated travel time is $L/25$ m/s, shown in red.

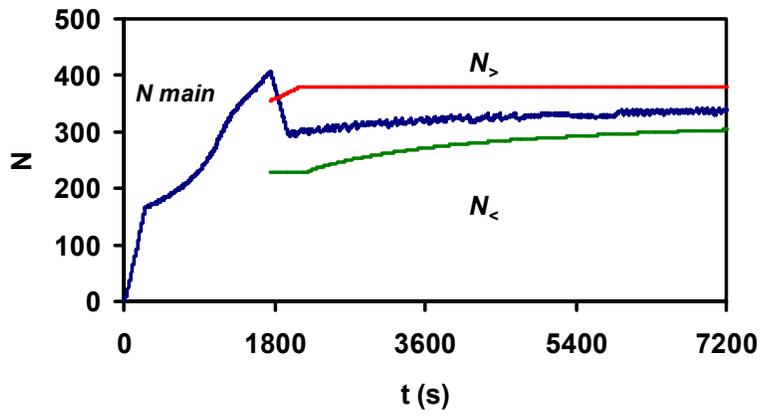

Fig. 23. The number of vehicles on the main line (blue curve) for the parameters of Figs. 21 and 22. The red and green curves are the averages $N_>$ and $N_<$, respectively.

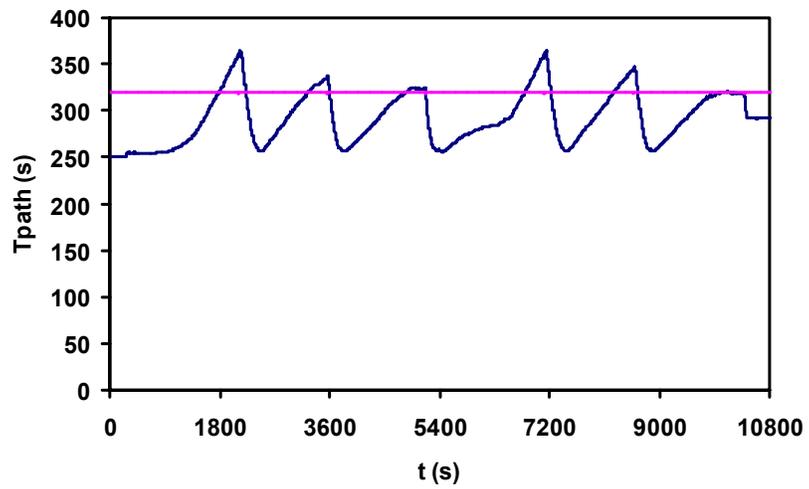

Fig. 24. The travel time as a function of time for the main line (blue) and bypass (red) for the parameters of Figs. 21 and 22. Drivers choose routes on the basis of selfish routing comparing $T_{path}^{main}$ to $T_{path}^{bypass}$.